\documentclass[
reprint, 
superscriptaddress,
 amsmath,
 amssymb,
 aps,
]{revtex4-2}
\usepackage{subfigure}
\usepackage{hyperref}
\usepackage{graphicx}
\usepackage{xcolor}
\usepackage{dcolumn}
\usepackage{dsfont}
\usepackage{stackengine}
\usepackage{bm}

\usepackage{physics}
\usepackage{gensymb}
\usepackage{amsmath}
\usepackage{soul}

\begin{document}

\title{Generating a 4-photon Tetrahedron State: Towards Simultaneous Super-sensitivity to Non-commuting Rotations}

\author{Hugo Ferretti}
    \email{hu.ferretti@gmail.com}
    \affiliation{Department of Physics and Centre for Quantum Information \& Quantum Control,University of Toronto, Toronto, Ontario M5S 1A7, Canada}

\author{Y. Batuhan Yilmaz}
    \email{ybylmaz@physics.utoronto.ca}
    \affiliation{Department of Physics and Centre for Quantum Information \& Quantum Control,University of Toronto, Toronto, Ontario M5S 1A7, Canada}

\author{Kent Bonsma-Fisher}
    \affiliation{National Research Council of Canada, 100 Sussex Drive, Ottawa, Ontario K1N 5A2, Canada}

\author{Aaron Z. Goldberg}
    \affiliation{National Research Council of Canada, 100 Sussex Drive, Ottawa, Ontario K1N 5A2, Canada}
    \affiliation{Department of Physics, University of Ottawa,  25 Templeton Street, Ottawa, Ontario K1N 6N5, Canada}

\author{Noah Lupu-Gladstein}
    \affiliation{Department of Physics and Centre for Quantum Information \& Quantum Control,University of Toronto, Toronto, Ontario M5S 1A7, Canada}
    
\author{Arthur O. T. Pang}
    \affiliation{Department of Physics and Centre for Quantum Information \& Quantum Control,University of Toronto, Toronto, Ontario M5S 1A7, Canada}

\author{Lee A. Rozema}
    \affiliation{University of Vienna, Faculty of Physics, Vienna Center for Quantum Science and Technology (VCQ) \& Research Network Quantum Aspects of Space Time (TURIS), 1090 Vienna, Austria}
    
\author{Aephraim M. Steinberg}
    \affiliation{Department of Physics and Centre for Quantum Information \& Quantum Control,University of Toronto, Toronto, Ontario M5S 1A7, Canada}
    \affiliation{CIFAR, 661 University Ave., Toronto, Ontario M5G 1M1, Canada}

\begin{abstract}
It is often thought that the super-sensitivity of a quantum state to an observable comes at the cost of a decreased sensitivity to other non-commuting observables. For example, a squeezed state squeezed in position quadrature is super-sensitive to position displacements, but very insensitive to momentum displacements.
This misconception was cleared with the introduction of the compass state \cite{zurek_sub-planck_2001}, a quantum state equally super-sensitive to displacements in position and momentum. 
When looking at quantum states used to measure spin rotations, N00N states are known to be more advantageous than classical methods as long as they are aligned to the rotation axis. When considering the estimation of a rotation with unknown direction and amplitude, a certain class of states stands out with interesting properties. 
These states are equally sensitive to rotations around any axis, are second-order unpolarized, and can possess the rotational properties of platonic solids in particular dimensions. 
Importantly, these states are optimal for simultaneously estimating the three parameters describing a rotation.
In the asymptotic limit, estimating all $d$ parameters describing a transformation simultaneously rather than sequentially can lead to a reduction of the appropriately-weighted sum of the measured parameters’ variances by a factor of $d$.
We report the experimental creation and characterization of the lowest-dimensional such state, which we call the “tetrahedron state” due to its tetrahedral symmetry. 
This tetrahedron state is created in the symmetric subspace of four optical photons’ polarization in a single spatial and temporal mode, which behaves as a spin-2 particle. While imperfections due to the hardware limit the performance of our method, we argue that better technology can improve our method to the point of outperforming any other existing strategy in per-photon comparisons. 

\end{abstract}

\maketitle

\section{\label{sec:Intro}Introduction}
Quantum metrology~\cite{giovannetti_quantum-enhanced_2004, giovannetti_quantum_2006, giovannetti_advances_2011} employs fundamentally quantum systems as measurement probes to overcome the ``standard quantum limit"~\cite{caves_quantum-mechanical_1981, paris_quantum_2009, riedel_atom-chip-based_2010, ockeloen_quantum_2013, sewell_magnetic_2012}, a statistical bound on the precision of any measurement with uncorrelated particles.
Already, ideas from this field of research have been implemented in gravitational wave detectors~\cite{caves_quantum-mechanical_1981, aasi_enhanced_2013} and could become important in a range of precision measurement device such as magnetometers~\cite{wasilewski_quantum_2010, koschorreck_sub-projection-noise_2010, brask_improved_2015} and atomic clocks ~\cite{ludlow_optical_2015, schulte_prospects_2020}.

One of the classic problems used to demonstrate the advantage of quantum metrology is the measurement of a phase difference $\theta$ between the two arms $a$ and $b$ of a balanced interferometer. 
For the best measurement scheme using $N$ separable photons, the error in the estimate of $\theta$, $\delta \theta$, scales as $\delta\theta \propto 1/\sqrt{N}$. 
This is the ``standard quantum limit"~\cite{caves_quantum-mechanical_1981, paris_quantum_2009} in this problem, which is reflective of the measurement process being independent for each photon. 
On the other hand, the best general measurement strategy for this estimation is to send all the photons in a superposition of being all in path $a$ and all in path $b$. 
This highly entangled configuration is called the ``$N00N$ state"~\cite{sanders_quantum_1989, boto_quantum_2000, lee_quantum_2002, dowling_quantum_2008} and with it, one can in theory achieve the Heisenberg  scaling ~\cite{dowling_correlated_1998, paris_quantum_2009} of $\delta \theta \propto 1/N$, which is the best scaling allowed by quantum mechanics, although its performance is quickly ruined by loss~\cite{gilbert_use_2008}. Low photon number $N00N$ states have been realized in the laboratory multiple times~\cite{mitchell_super-resolving_2004, walther_broglie_2004, nagata_beating_2007, afek_high-noon_2010, israel_experimental_2012, israel_supersensitive_2014, monz_14-qubit_2011, gao_experimental_2010}.
It is often thought that super-sensitivity of this type comes at a price and that a quantum state super-sensitive to a given transformation is necessarily less sensitive to transformations generated by non-commuting operators. 
For example, a $N00N$ state where $N$ photons are in a superposition of all being in $+z$ eigenstate and all being in $-z$ eigenstates has stellar sensitivity to rotation around the $z$ axis. However, its sensitivity to rotation around the $x$ and $y$ axes is bounded by the ``shot-noise limit" of $1/\sqrt{N}$. The 2-photon $N00N$ state is an extreme case which is completely insensitive to rotations around the $y$ axis.
The idea of the trade-off in super-sensitivity is a misconception which was first cleared with the introduction of the compass state~\cite{zurek_sub-planck_2001, toscano_sub-planck_2006}, a state of the quantum harmonic oscillator equally super-sensitive to displacement in position $\hat{x}$ and momentum $\hat{p}$.

In this work, we experimentally generate a counterpart of the compass state in the polarization of a light beam, the ``tetrahedron state".
Unlike in the $N00N$ state example, we no longer consider the characterization of a single rotation, described by the action of the group U(1) and generated by the angular momentum operator $\hat{J}_z$, but instead we must look at general rotations described by the actions of the group SU(2), which can be generated by a triad of angular momentum operators $\{\hat{J}_x, \hat{J}_y, \hat{J}_z\}$.
The tetrahedron state exhibits super-sensitivity for rotations generated by any linear combination of these three operators, corresponding to rotations around any axis, despite the non-commutativity of these generators.
Beyond polarization rotations, SU(2) operations include the action of a generalized interferometer (with arbitrary transmission and reflection) on two modes of the electromagnetic field, the action of arbitrary magnetic fields on spins, reference-frame alignment, and many other processes. 
We will consider scenarios where the axis of the rotation is known but not necessarily aligned with the $z$-axis, as well as scenarios where the axis itself is also unknown. 
The main difference present in the latter is that the characterisation of such processes requires the estimation of a minimum of three parameters instead of a single one~\cite{szczykulska_multi-parameter_2016, goldberg_quantum-limited_2018}.
The multi-parameter quantum estimation has proven to be more intricate than its single-parameter counterpart \cite{paris_quantum_2009, humphreys_quantum_2013, sidhu_geometric_2020, demkowicz-dobrzanski_multi-parameter_2020}, with nuances such as parameter compatibility \cite{zhu_information_2015, heinosaari_invitation_2016, ragy_compatibility_2016}, and is important for many real-world problems \cite{albarelli_perspective_2020, polino_photonic_2020}.
This makes for a richer problem and requires us to look in different corners of the Hilbert space for suitable entangled probe states.

The tetrahedron state on which we report has several intriguing properties. It was initially studied as being the opposite of SU(2)-coherent states, earning the epithet ``anticoherent''~\cite{zimba_anticoherent_2006, baguette_anticoherence_2015, baguette_anticoherence_2017}, making it in some sense the most quantum state. Using the Hilbert-Schmidt norm as the distance measure between states, it is the furthest state from the closest SU(2)-coherent state of the same dimension. Therefore, it has been termed as the ``Queen of quantumness''~\cite{giraud_quantifying_2010}.
In the context of polarization, it manifests ``hidden polarization''~\cite{klyshko_multiphoton_1992}, as its classical (first-order) polarization properties are ignorant of higher-order polarization features \cite{de_la_hoz_unpolarized_2014}.
It also minimizes the cumulative multipoles of its polarization distribution, for which it has earned the moniker ``King of quantumness"~\cite{bjork_stars_2015, bjork_extremal_2015}, and is the optimal state for rotation sensing, explaining the sobriquet ``Quantum rotosensor''~\cite{chryssomalakos_optimal_2017, martin_optimal_2020}. The tetrahedron state has numerous other extremal properties~\cite{goldberg_extremal_2020}, including maximizing delocalization measures in SU(2) phase space such as the Wehrl entropy \cite{baecklund_four_2014}. 
It has also been studied for its entanglement properties~\cite{aulbach_maximally_2010, martin_multiqubit_2010, ganczarek_barycentric_2012, burchardt_entanglement_2021, PhysRevA.105.022433}, which have been generalized to multi-dimensional systems with multiple qudits~\cite{chryssomalakos_symmetric_2021}. While its polarization-sensing properties require the tetrahedron state to be in light's polarization degree of freedom \cite{goldberg_quantum_2021}, it can also exist in light's spatial degrees of freedom \cite{bouchard_quantum_2017, gutierrez-cuevas_platonic_2020, eriksson_sensing_2023}, and can be envisioned in other symmetric multi-qubit systems.
We report on creating this unique quantum state in the polarization degree of freedom of light.

\section{\label{sec:theory}Theory}

\subsection{Quantum Fisher Information}
The Quantum Fisher Information (QFI)~\cite{helstrom_noncommuting_1974} is a powerful tool that, in the context of multi-parameter estimation, can provide a lower bound on the covariance matrix of the estimated parameters for any measurement scheme. 
Let $\bm{\tilde{\theta}}$ be the estimator for a set of parameters $\bm{\theta}$ describing a specific SU(2) rotation $\hat{R}(\bm{\theta})$; then, the covariance matrix
\begin{equation}
    \Sigma_{l,m} = E[(\theta_l -\tilde{\theta}_{l})(\theta_m -\tilde{\theta}_{m})] 
\end{equation}
is bounded from below by the quantum Cram\'er-Rao Bound (QCRB)~\cite{sidhu_geometric_2020}
\begin{equation}
    \Sigma \succeq I_{\rho, \bm{\theta}}^{-1},
    \label{eq:CRB}
\end{equation}
where $I_{\rho, \bm{\theta}}$ is the quantum Fisher Information matrix with components
\begin{equation}
    [I_{\rho, \bm{\theta}}]_{l,m} = \text{Tr}\left[\hat{\rho}_{\bm{\theta}}\left\{\hat{L}_{\theta_l},\hat{L}_{\theta_m}\right\}\right],
\end{equation} matrix inequalities $A\succeq B$ imply that $A-B$ is positive definite, and the expectation values $E[\cdot]$ are taken over the probability distribution for the data on which $\bm{\tilde{\theta}}$ is based.
The operator $\rho_{\bm{\theta}}$ is the density matrix $\hat{R}\left(\bm{\theta}\right)\rho\hat{R}\left(\bm{\theta}\right)^\dagger$ after it has been transformed by the action of the SU(2) rotation and the operators $\bm{\hat{L}}$ are the symmetric logarithmic derivative operators defined by 
\begin{equation}
    \partial_{\theta_l}\hat{\rho_{\bm{\theta}}} = \frac{1}{2} \left(\hat{\rho}_{\bm{\theta}}\hat{L}_{\theta_l} + \hat{L}_{\theta_l}\hat{\rho}_{\bm{\theta}}\right).
\end{equation}

To find the optimal state to characterize the SU(2) rotation, it is important to have a scalar bound to optimize. 
Equation~\ref{eq:CRB} can be easily turned into such a bound with the introduction of a cost matrix $C$. The cost function to optimise then becomes the weighted sum of the elements of the covariances
\begin{equation}
    \operatorname{Tr}[C \Sigma]\geq \operatorname{Tr}(C I_{\rho, \bm{\theta}}^{-1}) =: \mathcal{C}_{\rho, \bm{\theta}},
    \label{eq:scalar_bound}
\end{equation}
which is in turn bounded by $\mathcal{C}$, a scalar version of the QCRB (s-QCRB). Equation~\eqref{eq:scalar_bound} cannot always be saturated for multi-parameter estimation but can indeed be saturated using the tetrahedron state \cite{goldberg_rotation_2021}.
In general, the matrix $C$ can be arbitrary, providing different weightings to the different elements of the covariance matrix. 
When it comes to characterizing the elements of a group responsible for the transformation, a natural choice of the cost matrix is the Cartan metric of the group~\cite{helgason_differential_1979}.
With the cost metric so chosen, Equation~\ref{eq:scalar_bound} becomes completely independent of the parametrization of the group. 
It can be shown~\cite{goldberg_intrinsic_2021} that, for pure states $\ket{\psi}$ with this natural metric,
\begin{equation}
    \mathcal{C} = \frac{1}{4}\operatorname{Tr}[\operatorname{\bf{Cov}}^{-1}_{\psi}(\bm{\hat{J}})],
    \label{eq:coolbound}
\end{equation}
where $\bm{\hat{J}}$ are the generators of the Lie Algebra associated with the group and $\operatorname{\bf{Cov}}_\psi$ contains the symmetrized covariances of those operators, with matrix elements
\begin{equation}
    [\operatorname{\bf{Cov}}_{\psi}(\bm{\hat{J}})]_{l.m} =\frac{1}{2} \left\langle\hat{J}_l\hat{J}_m+\hat{J}_m\hat{J}_l\right\rangle - \left\langle\hat{J}_l\right\rangle\left\langle\hat{J}_m\right\rangle
\end{equation} and expectation values $\langle \cdot\rangle$ taken with respect to $|\psi\rangle$.

 It is noteworthy that the bound in Equation~\ref{eq:coolbound} is independent of both the parameters to estimate $\bm{\theta}$ and of the initial rotation of the state (i.e. $\ket{\psi}$ and $\hat{R}\left(\bm{\phi}\right)\ket{\psi}$ will share the same bound for all $\bm{\phi}$). 
 Furthermore, if we choose a Cartesian parametrization for the rotation
 \begin{equation}
     \hat{R}(\bm{\theta}) = e^{-i\bm{\theta}\cdot\bm{\hat{J}}},
 \end{equation}
 then the s-QCRB with the Cartan metric in Eq.~\ref{eq:coolbound} coincides with the s-QCRB with $C = \mathds{1}$ when the rotation is in the neighborhood of identity. 
 Then, $\mathcal{C}$ is a lower bound for $\sum_i \operatorname{Var}\left[\theta_i\right]$, the identically weighted sum of the variances of the three rotation angles.
 
\subsection{Polarization as a Spin}
A single photon's transverse polarization can be represented as a spin-$1/2$ particle, with the polarization states $\ket{H} (m=1/2)$ and $\ket{V} (m=-1/2)$ forming an orthonormal basis for the Hilbert space. 
In general, a collection of $N$ photons can be decomposed into the tensor sum of particles with spin $\leq N/2$: 
\begin{equation}
    \left(\text{spin-}1/2\right)^{\bigotimes N} = \bigoplus_{j = [0, 1/2]}^{N/2}d_j\left(\text{spin-}j\right),
\end{equation} 
where the coefficients $d_j$ represent the multiplicities of the different spin sectors. 
The highest-dimensional spin sector is the $\text{spin-}N/2$ and is unique ($d_{N/2}=1$). 
It represents the polarization states that are full-symmetric under the pairwise exchange of any particles. 
If there are no other degrees of freedom to distinguish photons, the bosonic nature of the photons would constrain their polarization state to this sector.
We can therefore identify the basis element of the polarization state of a $N$ photon in a single mode simply by $n_H$ and $n_V$, the number of H and V photons it contains.
For example, for a three-photon state, the following are equivalent, 
\begin{equation}
    \begin{split}
        &\ket{j = 3/2, m = 1/2} = \ket{2_H, 1_V} \\&= \frac{1}{\sqrt{3}} \left(\ket{HHV} + \ket{HVH} + \ket{VHH}\right).
    \end{split}
\end{equation}
One way to represent the generators of the SU(2) Lie algebra in this space is 
\begin{equation}
    \bm{\hat{J}} = \bigoplus_{j= [0,1/2]}^{N/2}
    \underbrace{\bm{\hat{J}}^{(j)}\oplus\cdots\oplus \bm{\hat{J}}^{(j)}}_{d_j\,\mathrm{times}}
\end{equation}
where $\bm{\hat{J}}^{(j)} = \left(J_x^{(j)},J_y^{(j)},J_z^{(j)}\right)$ are the usual spin angular momentum operator in a $\text{spin-}j$ space.
It is easy to prove~\cite{kolenderski_optimal_2008} that in this Hilbert space, the state minimizing $\mathcal{C}$ is contained in the full-symmetric $\text{spin-}N/2$ subspace.

\subsection{The Majorana Representation}

A helpful way to visualize states in the full-symmetric $\text{spin-}N/2$ sector is with the Majorana representation \cite{majorana_atomi_1932, bengtsson_geometry_2017}. 
The idea is that one can always write these states as a product of single-particle creation operators over the two polarization modes acting on the vacuum state:
\begin{equation}
\begin{aligned}
\sum_{m=0}^N\psi_m\ket{m_H,\left(N-m\right)_V}
\propto\prod_{k=1}^N \hat{a}_{\theta_k,\phi_k}^\dagger\ket{0_H,0_V}\\
\hat{a}_{\theta,\phi}^\dagger =\cos\frac{\theta}{2}\hat{a}_H^\dagger+e^{i\phi}\sin\frac{\theta}{2}\hat{a}_V^\dagger.
\end{aligned}
\end{equation}
When considered individually, these creation operators $\hat{a}_{\theta,\phi}^\dagger$ each produce a qubit state when acting on the vacuum. Moreover, this overall state is equivalent to the symmetric superposition of the tensor product of all $N$ such qubit states.
Hence, one can represent full-symmetric states of the polarization of $N$ photons by $N$ points on the Bloch sphere labeled by their angular coordinates $(\theta_k,\phi_k)$. 
The action of a SU(2) rotation is simply to rigidly rotate the representation by the corresponding 3D rotation of the sphere, which makes this visualization an integral tool to gain intuition about the action of rotation on different states.

\subsection{\label{sec:strategies}Choosing the Optimal State for Multiparameter Estimation}
\subsubsection{The spin coherent states}
Spin coherent states, depicted in Fig.~\ref{fig:concept}\,(a), are notable separable states in the symmetric subspace of $N$ photons. 
They consist of all $N$ photons sharing the same polarization properties. 
One of those states is 
\begin{equation}
    \ket{\psi_{coh}} = \ket{N_H, 0_V},
\end{equation}
where all the photons are horizontally polarized, and the rest can be obtained by rotating that state. 
The spin covariance matrix for this state is given by 
\begin{equation}
    \operatorname{\bf{Cov}}_{\psi_{coh}}(\bm{\hat{J}}) = \begin{pmatrix}
    N/4&0&0\\
    0&N/4&0\\
    0&0&0
    \end{pmatrix},
\end{equation}
and is not invertible, which leads to a divergence in the bound s-QCRB and an infinitely bad performance at fully characterizing SU(2) rotations. 
Intuitively, this is due to the state being completely undisturbed by any rotation around the polarization axis of each of the separable particles.
It is therefore incapable of estimating one of the three parameters necessary to characterize the rotation, which leads to a divergence in the weighted sum of the variance of those parameters. 
In the example above, all the polarizations of all photons were aligned with H, which leads to a variance of $0$ for $\hat{J}_z$, the spin operator aligned with that axis. 

To remedy this situation, we could imagine dividing our available photons into three different spin-coherent states aligned with the $x$, $y$ or $z$ axes and using them sequentially. 
Because the quantum Fisher information is additive for uncorrelated measurements~\cite{sidhu_geometric_2020}, the bound in Equation~\ref{eq:coolbound} can be found adding up the different spin covariance matrices for each of these states. 
Assuming $N$ is divisible by $3$, the resulting matrix and the corresponding s-QCRB are
\begin{align}
    \sum_i\operatorname{\bf{Cov}}_{\psi_{coh}^{(i)}}(\bm{\hat{J}}) = \frac{N}{6} \times \mathds{1},
    &&
    \mathcal{C} = \frac{9}{2N},
\end{align}
where $\mathds{1}$ is the $3\times3$ identity matrix.
As we can see, by splitting our photons into three batches, we retain sensitivity to every parameter and recover the expected shot-noise scaling from a classical state.
\subsubsection{The N00N states}
The $N00N$ state is an entangled state of $N$ photons. The particular N00N state aligned with the $z$ axis can be written as
\begin{equation}
    \ket{\psi_{N00N}} = \frac{1}{\sqrt{2}}(\ket{N_H,0_V} + \ket{0_H, N_V})
\end{equation}
As mentioned previously, it is notable for being the most sensitive state for measuring the phase difference between two arms of an interferometer, or equivalently for estimating the angle of a rotation around a known axis. The $N00N$ state aligned as above, for example, would be especially well suited to estimate the angle of a rotation around the $z$-axis. When it comes to estimating all parameters of a rotation, however, we see from the following spin covariance matrix and its corresponding bound

\begin{equation}
\begin{split}
    \operatorname{\bf{Cov}}_{\psi_{N00N}}(\bm{\hat{J}}) = \begin{pmatrix}
    N/4&0&0\\
    0&N/4&0\\
    0&0&N^2/4
    \end{pmatrix},
    \\
    \mathcal{C} = \frac{2}{N} + \frac{1}{N^2},
\end{split}
\end{equation}

that the performance of the $N00N$ state still exhibits a classical scaling. This is because the $N00N$ state fails at performing the estimation of the other two parameters with a quantum advantage and their contributions dominate in the high $N$ limit. 

We can attain a quantum scaling, however, by splitting the $N00N$ state into three batches, as with the spin coherent state. If the photons are split equally between $N00N$ states aligned with the $x$, $y$ and $z$ axes, we get
\begin{equation}
\begin{split}
    \sum_i\operatorname{\bf{Cov}}_{\psi_{N00N}^{(i)}}(\bm{\hat{J}}) = \frac{1}{36}\left(N\left(N+6\right)\right) \times \mathds{1},
    \\
    \mathcal{C} = \frac{27}{N(N+6)}.
\end{split}
\end{equation}
While splitting photons into three batches recovers the advantageous Heisenberg scaling, it causes us to pick up a factor of 3 in the leading order of $N$. Informally, we can think of the scaling as $\mathcal{O}(3 \times 1/ (N/3)^2) = \mathcal{O}(3^3/N^2)$.

\subsubsection{The second-order unpolarized states}

It was proven in~\cite{goldberg_extremal_2020, goldberg_rotation_2021} that the optimal states for the task at hand are pure states that exhibit the following properties:
\begin{align}
\expval{\bm{\hat{J}}} = 0 && \expval{\hat{J}_l\hat{J}_m} =\frac{N}{6}\left(\frac{N}{2} + 1\right) \delta_{l,m}.
\label{eq:second-order unpolarized defn}
\end{align}
These states have the property that they are isotropic up to the second moment of the spin angular-momentum operators, hence their name.  
Indeed, no initial rotation would change the two properties above for any state that satisfies them.
These states do not exist in all dimensions. 
It is, for example, impossible to satisfy this condition with one, two, or three photons.
There is one solution for four photons but then again none for five photons. 
The dimensionality of the spaces for which these states exist still constitutes a topic of research~\cite{baguette_anticoherence_2017, noauthor_extremal_nodate} that may be intimately connected to spherical $t$-designs~\cite{hardin_mclarens_1996, crann_spherical_2010, bannai_note_2011, bjork_stars_2015} and other problems for distributing points on the surface of a sphere~\cite{thomson_xxiv_1904, tammes_origin_1930, whyte_unique_1952, conway_packing_1996, saff_distributing_1997}. 
Multiple names mentioned above have already been used to describe these states and their usefulness has been theoretically demonstrated for a range of applications.

We can quickly see that states satisfying Eq.~\ref{eq:second-order unpolarized defn} outperform all the other schemes imagined here by looking at their spin covariance matrix and the s-QCRB:
\begin{equation}
\begin{split}
     \operatorname{\bf{Cov}}_{\psi_{plat}}(\bm{\hat{J}}) = \frac{1}{12}\left(N\left(N+2\right)\right) \times \mathds{1},
    \\
    \mathcal{C} = \frac{9}{N(N+2)}.
\end{split}
\end{equation}
The first thing to notice is that, in the leading order of $N$, these states recover the factor of $3$ that was lost from splitting the $N00N$ state photons into three batches. 
Indeed, our cost function is now bounded by something on the order of $\mathcal{O}(9/N^2)$ instead of $\mathcal{O}(27/N^2)$. 
This is in line with many recent studies on comparing simultaneous multi-parameter estimation and sequential parameter estimation in the quantum regime. 
It has been found that optimal simultaneous schemes often outperform sequential schemes by a factor of the number of parameters in the large N limit~\cite{humphreys_quantum_2013, goldberg_intrinsic_2021}. 

This is precisely $3$ in the case of the SU(2) rotation. The advantageous rotation-sensing performance of states with these properties has recently been demonstrated using light's orbital angular momentum degree of freedom, where operations that are mathematically equivalent to rotations are applied using multi-plane light conversion \cite{eriksson_sensing_2023}.

\subsection{The Tetrahedron State}
\begin{figure}[t]
    \centering
    \includegraphics[width=\linewidth]{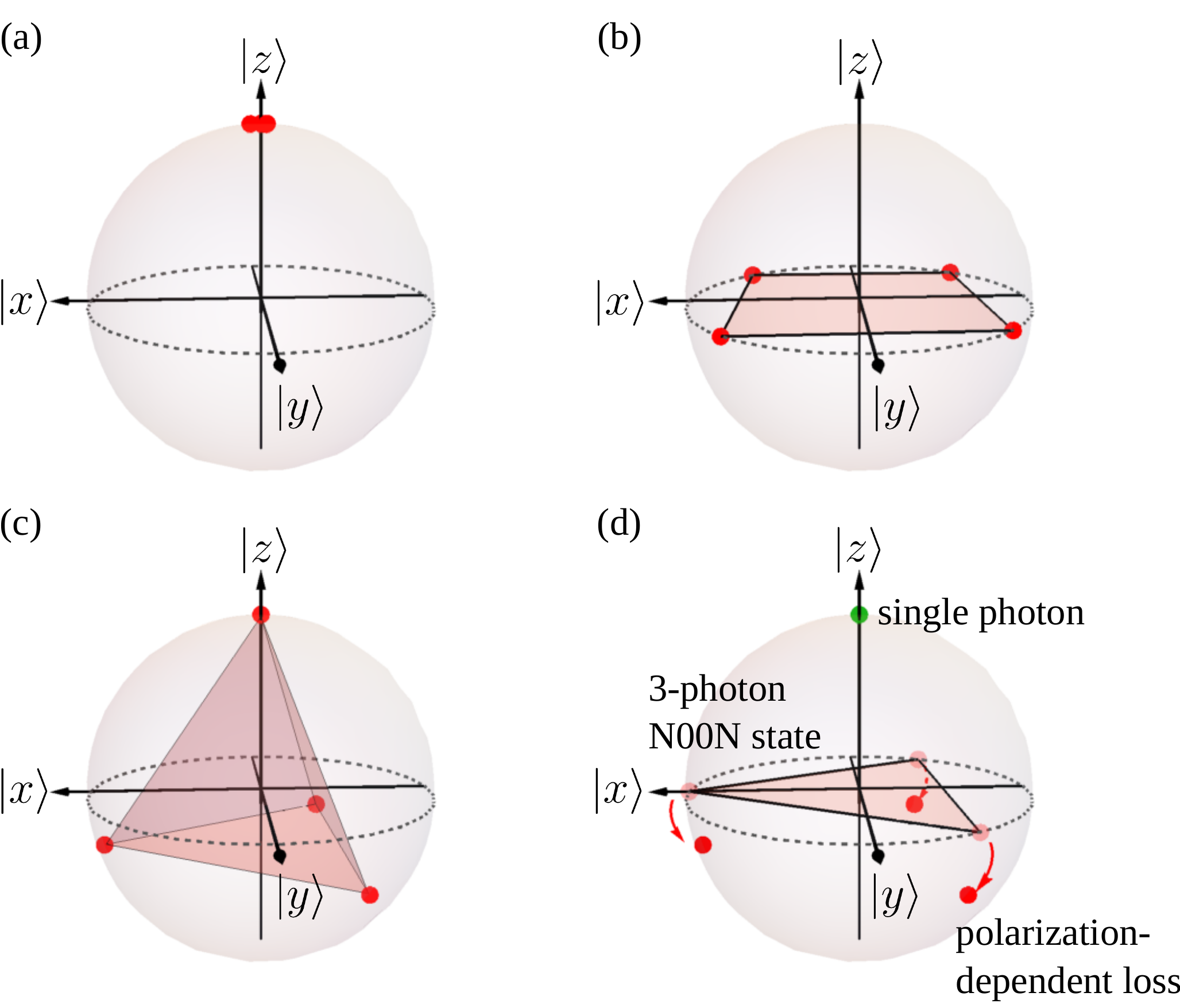}
    \caption{ Majorana representations of  (a) spin coherent state, (b) 4-photon $N00N$ state, and (c) the tetrahedron state. (d) Conceptual illustration of the tetrahedron's state creation process. The H polarization of a 3-photon $N00N$ state (three transparent red points on the vertices of an equilateral triangle on the equator) is partially attenuated resulting in the triangle of points with lower latitude (red). Simultaneously, an H polarized heralded single-photon is probabilistically added (green). The result is a 4-photon state represented by four points on the vertices of a tetrahedron.}
    \label{fig:concept}
\end{figure}

The second-order unpolarized state with the smallest photon number ($N=4$) is the one we call the ``tetrahedron state." 
It is named as such because it has the same symmetries as the tetrahedron under rotation, which is apparent in its Majorana representation (depicted in Fig.~\ref{fig:concept}(c)), where the points correspond to the vertices of a regular tetrahedron inscribed in the Bloch sphere.
The state can be written as 
\begin{equation}
    \ket{\psi_{tetra}}= \sqrt{\frac{1}{3}}\ket{4_H, 0_V} + \sqrt{\frac{2}{3}}\ket{1_H, 3_V}
\end{equation}
In this work, we create the tetrahedron state in the polarization state of four photons in the same spatial and temporal mode, which ideally guarantees the polarization to be full-symmetric.

\section{\label{sec:results}Methods and Results}

\subsection{Experimental Apparatus}

Our overall approach to our tetrahedron state creation scheme is intuitive in the Majorana picture. When we rewrite the tetrahedron state as a product of creation operators acting on the vacuum and we re-arrange the terms we get the following expression:
\begin{equation}
    \ket{\psi_{tetra}} = \frac{1}{3}\left(\left(\frac{\hat{a}_H^\dagger}{\sqrt{2}}\right)^3 + \hat{a}_V^{\dagger^3}\right)\hat{a}_H^\dagger\ket{0}.
\end{equation}

As can be seen, the tetrahedron consists of a combination of a three-photon N00N state in the H/V basis for which the H polarization has been attenuated by a factor of $2$\, and a horizontally polarized single photon\, respectively.

\begin{figure*}[t]
    \centering
    \includegraphics[width=0.9\linewidth]{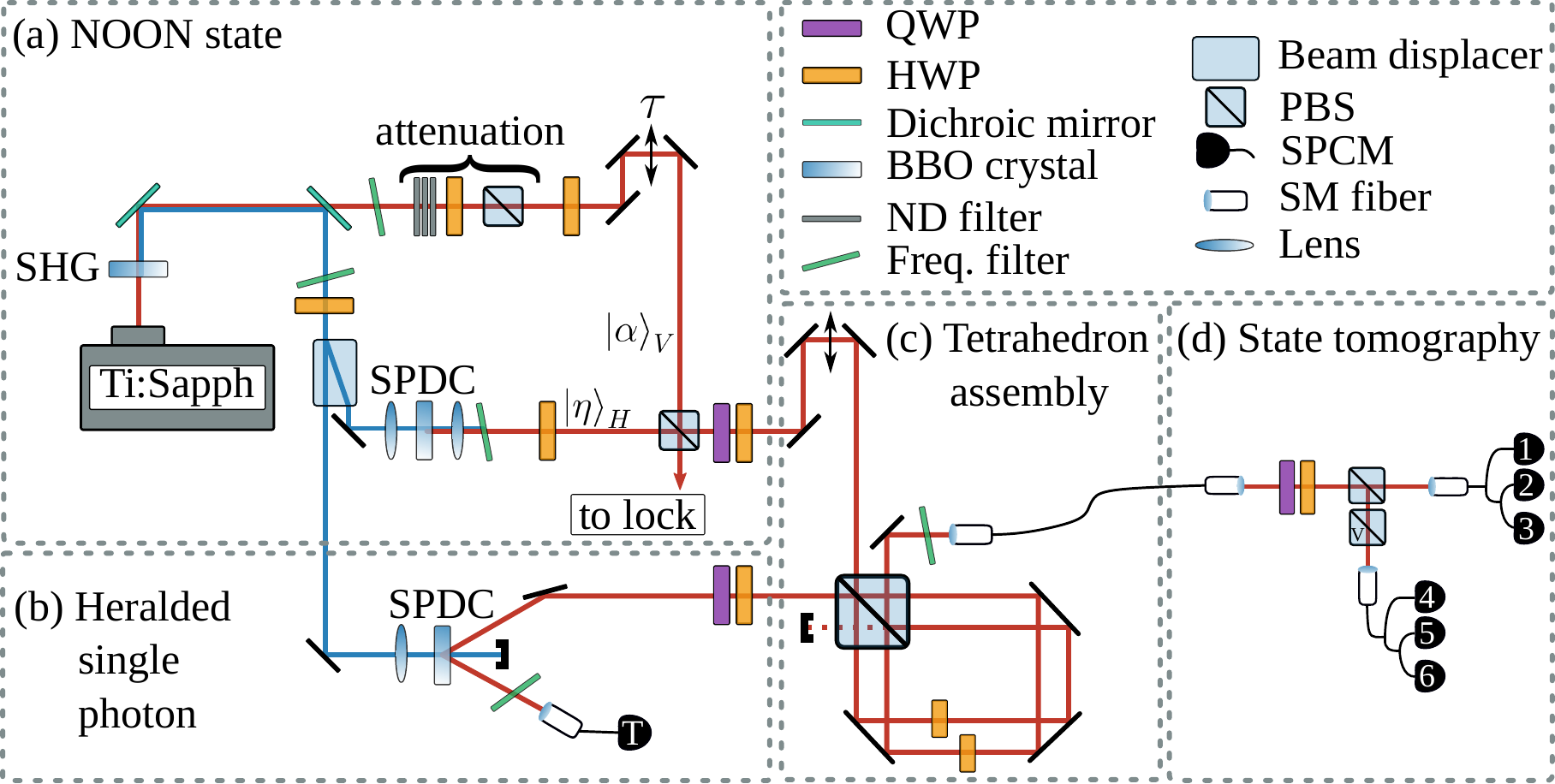}
    \caption{Experiment overview. (a) An 808\,nm pulsed beam (red) is used to generate a 404\,nm beam (blue) through second harmonic generation (SHG) in a 1\,mm barium borate (BBO) crystal. Part of this blue beam is sent to generate co-propagating pairs of horizontally polarized 808\,nm photons (red) through type-I spontaneous parametric down-conversion (SPDC) in a 2\,mm BBO crystal while the red beam is attenuated and vertically polarized. The two beams are recombined on a polarizing beamsplitter (PBS) and polarization rotated to form a three-photon $N00N$ state in the H/V basis. (b) Another blue beam then a pumps a second SPDC (2\,mm BBO) to generate another photon pair. Here, after a 3\,nm narrow-band filter, a detection at a single-photon counting module (SPCM) labelled ``T" heralds the presence of a photon in the upper path. (c) Both the $N00N$ state and heralded single photon enter a displaced polarization Sagnac interferometer. The heralded photon strictly takes the clockwise path while the $N00N$ state is split equally amongst the paths. The half-waveplate (HWP) in the counter-clockwise path is set at $45 ^{\circ}$ to transmit all the light in the upwards output while the HWP in the clockwise path is set at $22.5 ^{\circ}$ to attenuate the V polarization of the $N00N$ state by half and transmit the heralded single-photon half the time (see Figure \ref{fig:concept} for context). After being recombined, all the photons pass through a 3\,nm narrow-band filter to make sure their spectra are the same. (d) The state is transmitted via single-mode (SM) fiber to a tomography apparatus for projection onto an arbitrary polarization axis. Each output path is coupled to a multi-mode fiber splitter to allow for partial photon number counting capabilities (a maximum of three detections per polarization). Five-fold detection events between T and any four SPCMs in the tomography setup are recorded.}
    \label{fig:setup}
\end{figure*}

We create these two components independently with SPDC. The $N00N$ state is created by an interference between a weak single-mode squeezed vacuum generated through SPDC, and a coherent state \cite{hofmann_high-photon-number_2007, afek_high-noon_2010, rozema_optimizing_2014}. The relevant section experimental apparatus is depicted in Figure~\ref{fig:setup}(a). We use a pulsed Ti-Saph laser with a pulse width of 140\, fs, a center wavelength of  807\,nm, a repetition rate of $80$\,MHz, and an average intensity of $3.5$\, W. With it, we perform SHG in a $1$\,mm BBO crystal to get a beam of over $500$\,mW at $404$\, nm. This beam, blue in Fig.\,\ref{fig:setup}, serves as a pump for both our SPDC processes. In the N00N state source, we perform SPDC with a $2$\,mm BBO crystal. This SPDC source is optimized in Type 1 collinear configuration. We measure a pair creation rate of $53$ kHz with a symmetric coupling efficiency of $13.4\%$. These numbers include all forms of loss in the experiment, including detector inefficiencies, but are corrected to account for the necessity to probabilistically split the photons in a collinear geometry. These photon pairs interfere with a coherent state in a polarisation interferometer with a Mach-Zehnder geometry. The coherent state is light from the attenuated original laser beam, left over from the SHG process. 

Importantly, the phase between these two beams should be stable throughout the experiment. We actively stabilize this phase using the locking port in Figure\,\ref{fig:setup}(a). The squeezed vacuum and the coherent states are minutely off from their intended polarizations so that we can collect a bit of light from the locking port. The superposition of 2 horizontally polarized single photons from the squeezed vacuum and 2 vertically polarized single photons from the coherent state forms a 2-photon N00N state since those 2 terms are mostly indistinguishable in any other degrees of freedom. We measure 2-photon coincidence events on a basis where the correct phase between two terms would result in a signal that is in the middle of a fringe. Because of the drifts in the rates, we re-calibrate the set point every 8-12 hours of data collection.

The horizontally polarized single photon, is also created through SPDC, and depicted in Figure\,\ref{fig:setup}(c). This time we use a non-colinear Type I geometry. One of the photons of the pair is detected immediately and acts as a herald, while the other goes to be part of the tetrahedron state. Throughout this experiment, this SPDC source produced photon pairs at a rate of $32$\,kHz and with a heralding efficiency of $11.5\%$. Using a heralded single photon here as opposed to a low-intensity coherent state was necessary to suppress the contamination from higher order terms where more than one photon would be created in this mode. In order to ensure that the photons from these different sources are as indistinguishable as possible, all the photons in the tetrahedron state are fed into the same single-mode fiber right after transmission through a $3$\, nm narrow-band filter. The resulting loss is already included in the efficiency figures above. The herald photon is similarly filtered and collected. 

As we will discuss in Section \ref{sec:discussion}, the state we generated is notably far from ideal. The main source of imperfections is the contributions of the higher-order terms from the coherent state and the SPDC sources. The four-photon tetrahedron state is created in a single spatial mode and a temporal mode coincidentally with a single ``herald" photon in another spatial mode. It is only when we detect these five photons simultaneously that we consider that a tetrahedron state has been created and detected. Ideally, in the absence of photon loss of any kind, the number of detected photons is the same as the number of created photons. In a laboratory setting, however, photon loss is unavoidable. As a result, some events with successful postselection have more than the five intended photons and some photons are lost. The polarization state of four of those remaining photons is not in a tetrahedron state. For a quantitative comparison, we performed sets of measurements to find the values of $\eta$, $\mu$, the squeezing parameters of collinear and non-collinear SPDC sources respectively. For both sources, the coincidence rate would be proportional to the absolute value squared of these squeezing parameters. We also measure $t$ and $\tau$, the transmission rates for the collinear SPDC and non-collinear SPDC photons respectively. These include the detection efficiency. The coherent state amplitude, $\alpha$(the single photon rate is proportional to $|\alpha|^2$), is set to $\sqrt{2\eta t}$ to create the 3-photon N00N state\cite{hofmann_high-photon-number_2007}. The measured values for $\eta$, $\mu$, $t$, and $\tau$ are  0.078, 0.14, 0.16, and 0.12, respectively. Transmission rates are quite low and this causes the other undesired events to become non-negligible. 

We calculated analytical expressions for the probabilities of possible 5-photon events. For example, the probability of the desired 5-photon event that generates a tetrahedron state is proportional to $ \eta^3 t^3 \mu^2 \tau^2 $. One of the biggest sources of imperfection is measuring an extra photon from the coherent state used to produce the $N00N$ state when one of the photons from the SPDC sources is lost. In these cases, the probability of the case when one of the photons from the collinear SPDC is lost is proportional to $ \eta^4 t^3(1-t) \mu^2 \tau^2 $, and the probability of the case when the horizontally polarized photon from the non-collinear SPDC is lost is proportional to $ \eta^4 t^4 \mu^2 \tau (1-\tau) $. All the cases mentioned above result in a 5-photon coincidence and we do not have a method to distinguish them. Substituting the values for our experimental setup, the two undesired events mentioned above have probabilities roughly on the same order of magnitude compared to the probability of generating a tetrahedron state. Several undesired events like the two mentioned above reduce the fidelity and the metrological potential of our state. One possible solution to lower the contributions from the undesired terms was to lower the squeezing parameters and the coherent state amplitude by reducing the single-photon rates. However, the data collection process, explained in Section \ref{tomohidden}, already took 8 days even with the current single-photon rates because of the rarity of the 5-photon events and the necessity of regular calibrations every 8-12 hours. Due to these impracticalities, we decided to collect data with the current rates.

\subsection{Tomography with Hidden Differences}\label{tomohidden} 
The Hilbert space for 4 photons' polarization is $\left(\text{spin-}1/2\right)^{\bigotimes 4} = \left(\text{spin-}2\right) \bigoplus 3\left(\text{spin-}1\right)\bigoplus2\left(\text{spin-}0\right)$. 
Ideally, in this case, the photons are indistinguishable, which constrains their polarization to be full-symmetric and contained in the $\text{spin-}2$ sector. 
In practice, however, we can expect some small mismatch in the modes of the photons from different sources, which will allow other spin sectors to be populated. 
Importantly, the parasitic distinguishability between the photons cannot be accessed, i.e., we have no practical way of favorably selecting photons from one source over another, which means that the only measurements we can perform have projectors symmetric under particle exchange. 
As a result, only the parts of the density matrix depicted in Fig.~\ref{fig:DM}\,(a) can be reconstructed.
They specifically consist of the spin sectors in the tensor sum decomposition of the Hilbert space. 
The coherences between these sectors are inaccessible. 
Furthermore, we are only sensitive to the sum of the different spin sectors with the same spin value and choose to report their average in each of the different sectors. 
Importantly, although our knowledge of the full state of the system is fundamentally limited in this way, the accessible information is enough to predict the result of any symmetric measurement done on the same state. 

The technique we use to reconstruct what part we need of the density matrix is called ``Tomography with Hidden Differences"~\cite{adamson_multiparticle_2007, adamson_detecting_2008, shalm_squeezing_2009}. 
It consists of a sequence of measurements on the polarization of our state akin to Stern-Gerlach measurements for a spin. 
The relevant section of the experimental apparatus is depicted in Fig.~\ref{fig:setup}\,(d). 
We send photons through a PBS and then through networks of beamsplitters~(BS) with every output leading to a single-photon counting module\,(SPCM). 
When four photons are simultaneously detected in coincidence with the herald photon in Fig.~\ref{fig:setup}\,(b), we consider it a successful event and count the number of photons transmitted vs. reflected at the PBS. 
We perform this polarization measurement in 13 different bases, chosen to be roughly uniformly spread on the Bloch sphere
by rotating the quarter-waveplate\,(QWP) and HWP before the PBS. 
With only three SPCMs on each output of the PBS, we cannot detect two of the five potential 4-photon outcomes, when the photons are all transmitted, or all reflected. 
With 13 measurement bases, however, we can still perform 39 different projections, which is enough to estimate the 35 linearly independent parameters of the accessible section of the density matrix using a maximum likelihood method.

Overall, 2434 successful events were recorded in 85 hours of counting time.
These counting hours were distributed over the course of 8 days. Every 8-12 hours, minor realignments were made to readjust the couplings in single-mode fibers and to reset the set-point of the locked interferometer.

Figure~\ref{fig:DM} shows the tomographic reconstruction of the density matrix. 
For ease of comparison with the theoretical tetrahedron state, we rotate our reconstructed density matrix by an optimal angle $\phi$ around the H/V axis in post-processing. 
We set $\phi$ by maximising the fidelity ($F = \bra{\psi_{tetra}}e^{-i\hat{J}_z\phi}\rho e^{i\hat{J}_z\phi}\ket{\psi_{tetra}}$) between the unrotated reconstructed state and the tetrahedron state and find a value of $\phi = 0.135$ for a fidelity with the tetrahedron state of $(0.46 \pm 0.02)$.

Error bars on the all the entries of the density matrix were determined by a Monte-Carlo simulation of the tomographic reconstruction process, where the simulated number of detection events for each projection was drawn from a Poisson distribution centered on the actual measurement results.

\section{\label{sec:discussion}Discussion}

\begin{figure}[t]
    \centering
    \includegraphics[trim = 50 10 0 0, clip, width=\linewidth ]{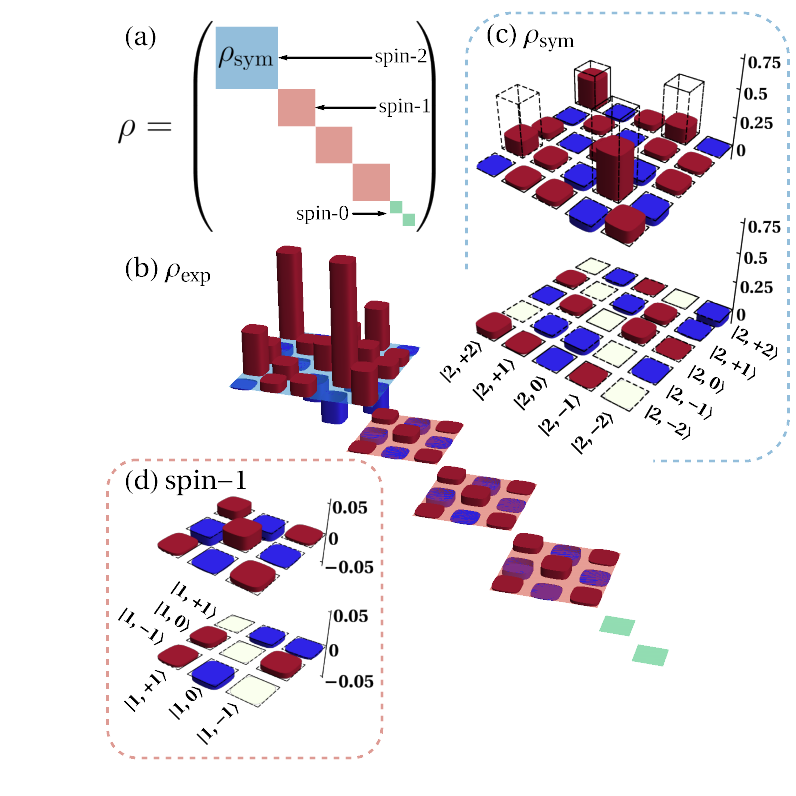}
    \caption{(a) Layout of the full density matrix. $\rho_\text{sym}$ is the fully symmetric spin-2 sector (blue). Spin-1 (red) and spin-0 (green) sectors are also shown. (b) The full tomographically reconstructed density matrix, $\rho_\text{exp}$. In hidden differences tomography, coherences between spin sectors are neglected and sectors of the same spins are chosen to be identical. (c) Real and imaginary parts of the reconstructed $\rho_\text{sym}$. The density matrix of the ideal tetrahedron state is shown for comparison (dashed). Elements of the density matrix are labeled with spin notation $\ket{j, m}$. (d) Real and imaginary parts of a spin-1 sector. The population in each spin-0 sector is less than $10^{-10}$.}
    \label{fig:DM}
\end{figure}

Despite the low fidelity between the theoretical tetrahedron state and our reconstructed density matrix, it retains many of the notable features of the former. 
As expected, most of the population ($87 \pm 4\%$) is in the full-symmetric, spin-2 subspace ($\rho_{\mathrm{sym}}$ in Fig.~\ref{fig:DM}\,(a)). The excess population in the other sectors is small but still non-zero, indicating that there were small mode mismatches between the different photons. 
As intended, the two basis elements with the highest population in $\rho_{\mathrm{sym}}$ are $\ket{2,2}$ and $\ket{2,-1}$. The ratio between these populations however is $(0.73 \pm 0.12)$, which is significantly larger than the theoretical value of $1/2$. 
We unfortunately measure significant population ($\approx 10\%$) in the states $\ket{2,0}$ and $\ket{2,-2}$ that are meant to be empty. 
Finally, the largest coherence measured is between $\ket{2,2}$ and $\ket{2,-1}$. Theoretically, this is the only non-zero coherence and it takes a value of $\sqrt{2}/3\approx 0.47$. 
Here, we measure the significantly lower value of $(0.13 \pm 0.02)$, also much lower than $(0.32 \pm 0.02)$, the maximum possible value given the corresponding population. 
The coherence quoted has no imaginary part specifically because we allowed an optimal rotation around the H/V axis in the reconstructed density matrix.
Most of these discrepancies can be explained by considering the contributions of terms created with six or more photons but detected as a successful five-fold coincidence due to photon loss at the detectors or in the optical apparatus. 
\begin{figure}[t]
    \centering
    \includegraphics[width=\linewidth]{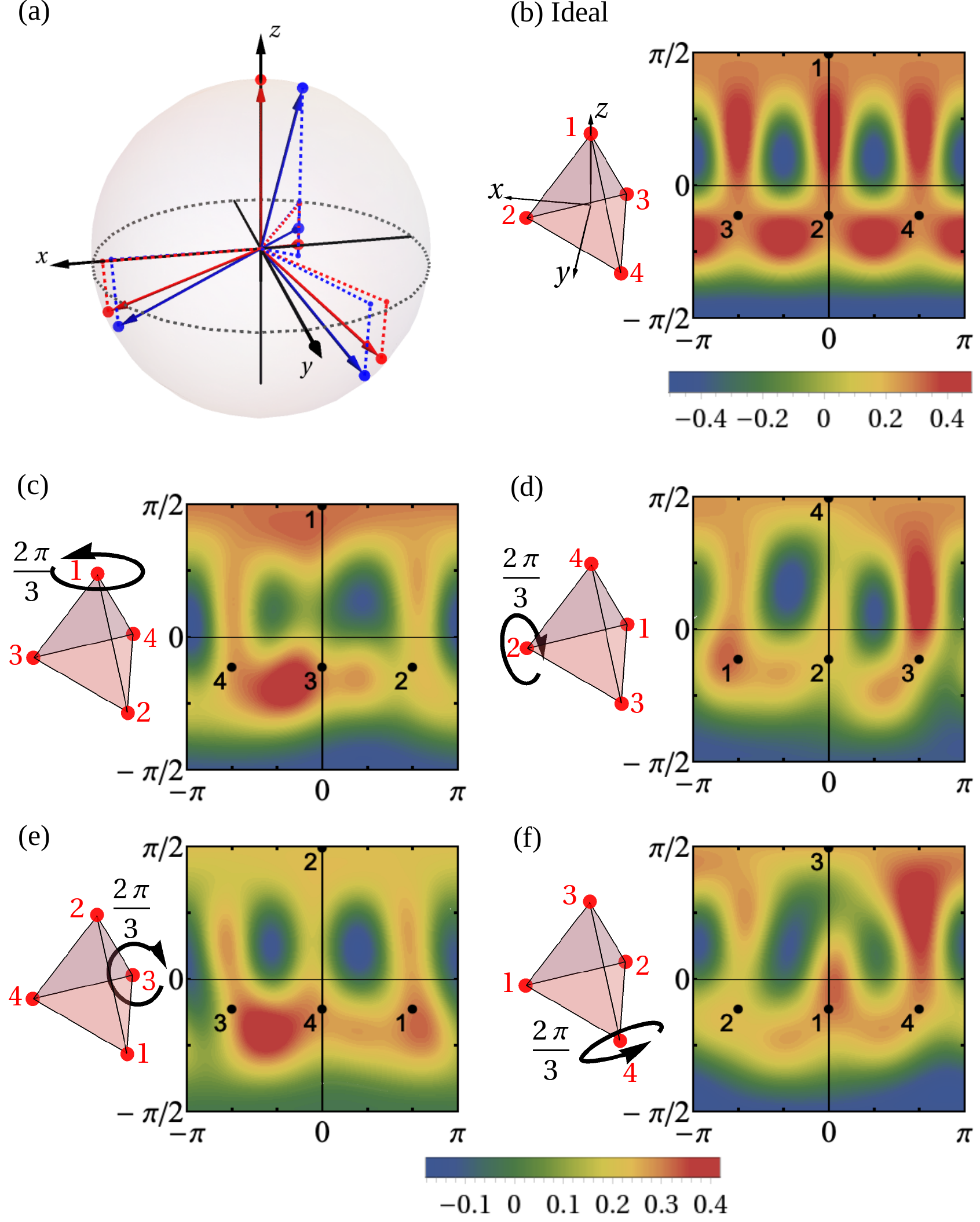}
    \caption{(a) Majorana representations for the ideal tetrahedron state~(red) and the eigenstate of the reconstructed density matrix with the largest eigenvalue~(blue). (b) Longitude-latitude map of the Wigner representation for the fully symmetric portion of the ideal tetrahedron state. Points on the tetrahedron representation correspond to those on the map. The y-axis is the polar angle and the x-axis is the azimuthal angle in spherical coordinates. Polar angle is defined to be $\pi /2$ at the top corner and azimuthal angle is defined to be 0 at the leftmost corner.  (c-f) The reconstructed Wigner representation when a $2\pi/3$ rotation is applied about each of the four points of the tetrahedron.}
    \label{fig:wigner}
\end{figure}

Despite these imperfections, the main eigenstate of the reconstructed density matrix (i.e. the state that was created most of the time, $49\%$ of the time), is very similar to the theoretical tetrahedron state with $92\%$ fidelity. We can interpret this to mean that our method succeeds in preparing a near-optimal tetrahedron state around half the time, but fails for the remaining half, in good part due to events when more photons were created than intended. 
A comparison of their Majorana representations is depicted in Fig.~\ref{fig:wigner}\,(a). 
Another feature of note is that our state keeps some of the tetrahedral symmetry that gives the tetrahedron state its name. 
This can be readily seen in the Wigner function~\cite{wigner_quantum_1932, dowling_wigner_1994, schmied_tomographic_2011} of our reconstructed state plotted in Figure~\ref{fig:wigner}. 
Because this Wigner function has the domain of a sphere, we display four different projections, each centered on a different vertex of a tetrahedron. 
These four different projections of the Wigner function of the theoretical tetrahedron state are the same (shown in Fig.~\ref{fig:wigner}\,(b)) by virtue of its symmetry. 
The different projections for the reconstructed experimental state are plotted in Fig.~\ref{fig:wigner}\,(c-f). 
Although, the features are not as pronounced, we notice that the troughs and the peaks of the Wigner function align with those of the theoretical state and that the reconstructed Wigner function qualitatively keeps, with some imperfections, the theoretical symmetry of the tetrahedron state. 

\begin{figure}[!ht]
    \begin{subfigure}
          \centering
    \includegraphics[width=0.9\linewidth]{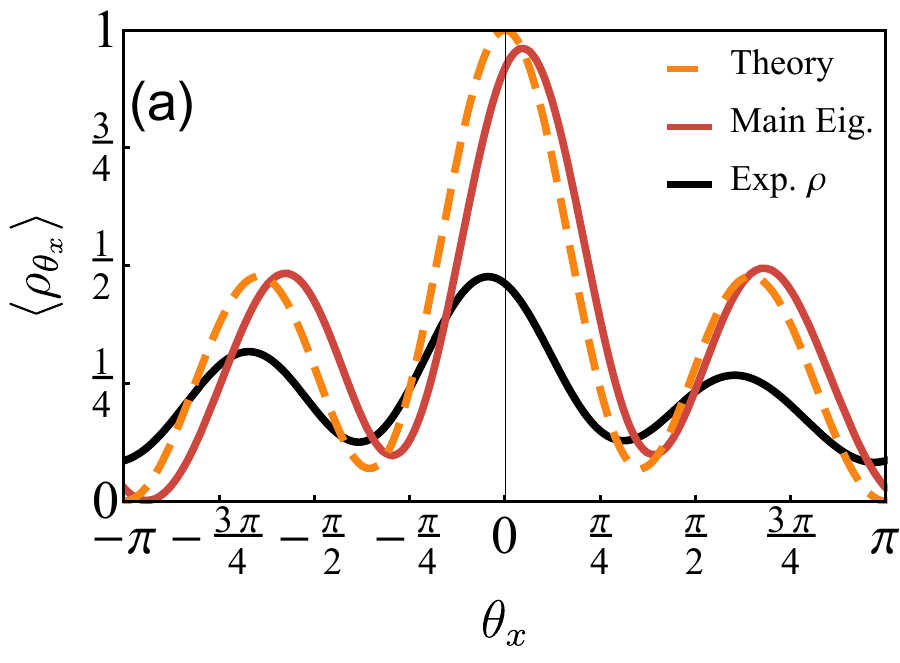}
    \end{subfigure}
     \begin{subfigure}
          \centering
    \includegraphics[width=0.9\linewidth]{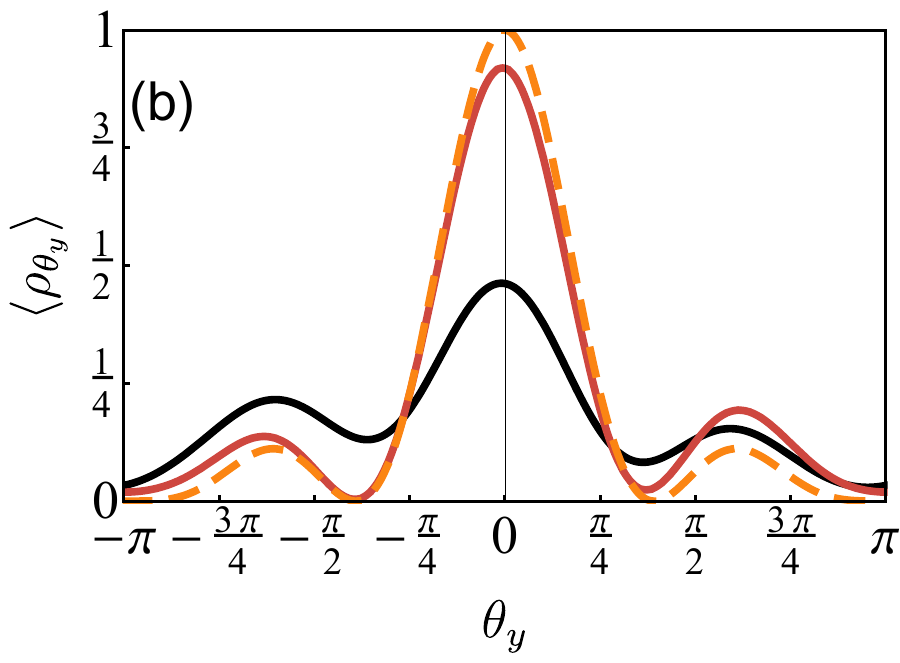}
    \end{subfigure}
      \begin{subfigure}
          \centering
    \includegraphics[width=0.9\linewidth]{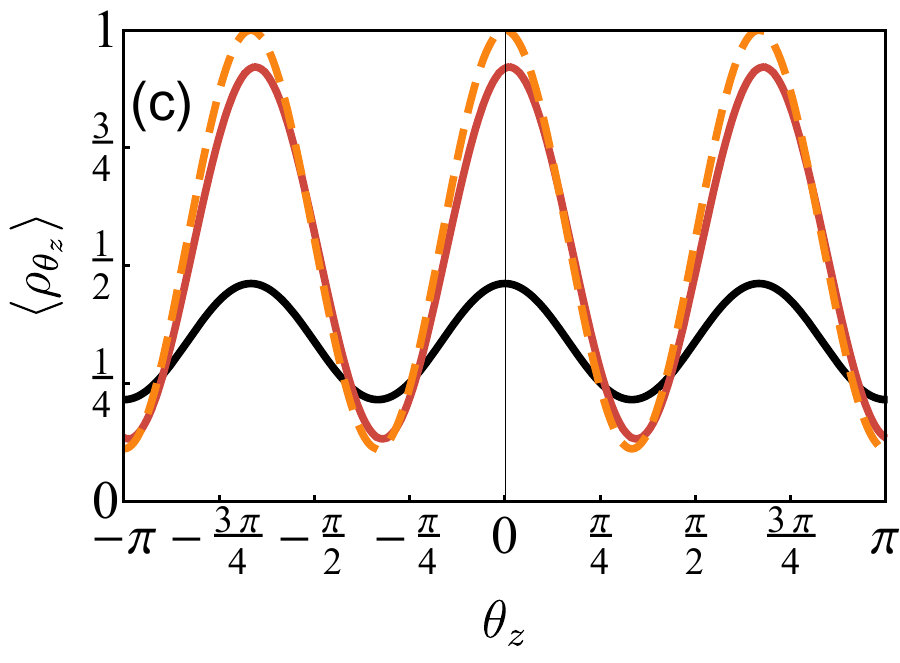}
    \end{subfigure}
    \caption{ Numerically calculated projection on the original tetrahedron state after rotations around x,y, and z axes on Bloch Sphere for the following states: Ideal tetrahedron state (dashed orange), experimental density matrix obtained from the tomography (solid black), and the main eigenstate of the experimental density matrix (solid red). The x-axis of the figure is the rotation amplitude in radians and the y-axis is the overlap between the rotated state and the ideal tetrahedron state.  }
    \label{fig:singleparameter}
\end{figure}

We can also directly assess how well our state satisfies the unpolarized properties of Eq. \eqref{eq:second-order unpolarized defn}. One direct measure of this property is by expanding the state's quasiprobability distribution on a sphere in terms of multipole moments. Unpolarized states have vanishing first and second moments. Our state's first and second moments are  $1.6\times 10^{-2}$, and $3.1\times 10^{-2}$, respectively. This is much smaller than for a comparable ``classical'' state, with first- and second-order multipole moments summing to $2/5$ and $2/7$, respectively, and is even much smaller than for a state whose Majorana constellation is chosen randomly \cite{goldberg_random_2021}, with first- and second-order moments summing to $2.8\times 10^{-1}$ and $2.3\times 10^{-1}$, respectively.

The qualitative features can be further observed in Figure~\ref{fig:singleparameter}, which compares the performance of the experimental state and the main eigenstate to the ideal tetrahedron state for the measurement of single rotations angles, that is the measurement of $\theta_x$, $\theta_y$ or $\theta_z$ when knowing that the other two angles take a value of 0. 
We plot the numerically calculated result of a projection on the original tetrahedron state following the rotation. 
It can be seen that both for the experimental state and the main eigenstate the visibility is not as high as the ideal tetrahedron state since they don't have perfect overlap with the tetrahedron state, especially the experimental state has very low visibility due to low fidelity and low purity. However, all axes still have three maxima per rotation, a signature of tetrahedral symmetry. 
For example, the 4-photon spin coherent state would have a single maximum, while the 4-photon $N00N$ state would exhibit 4 maxima for the z-axis, and two for the x-axis and the y-axis. The main experimental eigenstate closely 
approximates the tetrahedron state with similar levels of visibility for all rotations shown in the figure. 

\begin{figure}[t]
    \centering
  \includegraphics[width=\linewidth]{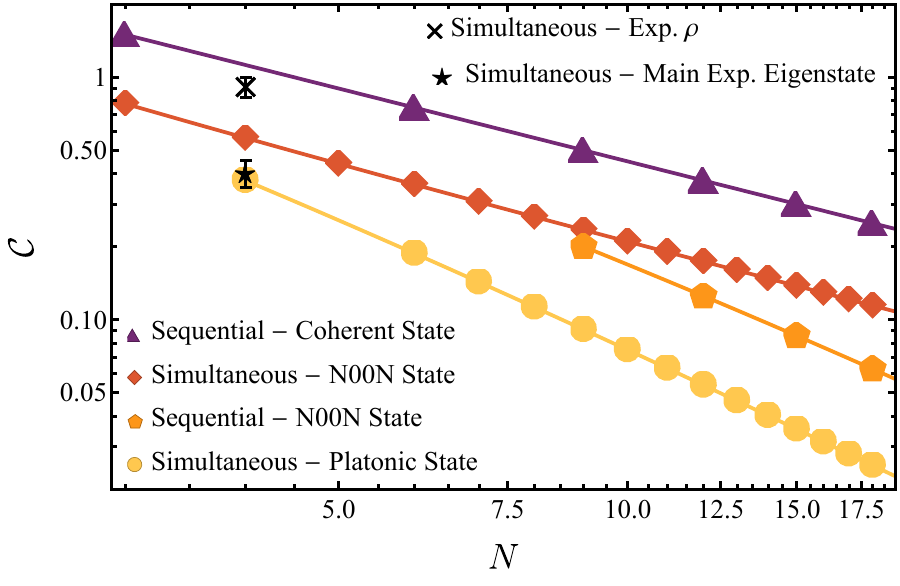}%
    \caption{Scalar quantum Cram\'er-Rao Bound $\mathcal{C}$ bound as a function of total number of photons. Solid lines represent the different strategies enumerated in Section \ref{sec:strategies}: spin-coherent states (purple triangles), N00N states (red squares for simultaneous, orange pentagons for sequential), and Platonic states (yellow circles). Sequential strategies use 3 batches of states aligned in the x, y, and z axes, whereas simultaneous strategies use only a single batch of identical states. The s-QCRB decreases with shot-noise scaling for sequential coherent states and simultaneous N00N states, and decreases with Heisenberg scaling for sequential N00N states and Platonic states. The s-QCRB of the experimentally reconstructed density matrix of the tetrahedron state is represented by a cross symbol. The black star represents the s-QCRB of its main eigenstate.}
    \label{fig:scalings}
\end{figure}

Finally, returning to multi-parameter estimation in Figure~\ref{fig:scalings}, we compare the s-QCRB calculated for our experimental state to the theoretical estimation strategies outlined in Section
\ref{sec:strategies}. As mentioned in that section, spin-coherent states have shot-noise scaling. Although $N00N$ states have Heisenberg scaling for one axis, other axes have shot-noise scaling and they dominate the resulting s-QCRB. To obtain Heisenberg scaling with $N00N$ states, one has to use 3 batches aligned in the $x$, $y$, and $z$ axes sequentially. Since one does not need a sequential strategy for Platonic states, they yield an s-QCRB 3 times better than sequential $N00N$ states. The figure has two data points, one for the experimental density matrix reconstructed from the tomography, and one for the main eigenstate of the density matrix. The s-QCRB for them are calculated theoretically from the reconstructed density matrix and the error bars are obtained using Monte-Carlo simulations. Having an s-QCRB of $0.91 \pm 0.09$, it narrowly beats the s-QCRB of the 4-photon spin-coherent state, which illustrates that the current approach to the creation of this particular state needs to be improved to be used in a real metrological task. 
The main experimental eigenstate comes very close to the ideal s-QCRB of $0.375$, beating all other outlined strategies for this photon number with an s-QCRB of $0.40 \pm 0.05$, demonstrating the potential of our method. 

Improvements in existing technology can render the creation of platonic states more feasible. In particular, as we discussed, multi-photon emissions are one of the main reasons for our low fidelity.  Another experimental imperfection arises from our use of partially distinguishable SPDC photons, which reduces the Hong-Ou-Mandel visibility both in our N00N state and tetrahedron assembly  stages.  To combat this we use narrowband filters in our experiment. However, this comes at the cost of decreasing our coupling efficiency; thus, increasing the multi-photon contamination. These challenges could be simultaneously overcome by the use of modern quantum dot single photon sources~\cite{tomm2021bright, wang2019towards}.  Such sources have recently been used for multi-photon interference experiments with up to 20 photons~\cite{PhysRevLett.123.250503}. With these on-demand photon sources one can imagine building the platonic states using the methods of references~\cite{mitchell_super-resolving_2004,shalm_squeezing_2009, PhysRevA.65.053818}. In this approach, independent photons are nondeterministically combined into the same spatial mode one at a time, each with a polarization state corresponding to a Majorana point. Although the state-creation methods of \cite{PhysRevA.65.053818} are probabilistic, given the high degree of indistinguishability and almost-zero multi-photon emission rates of quantum dot sources, the post-selected multi-photon states created from this method should not suffer the low fidelity we face here. Thus as quantum technology continues to improve it should be possible to harness the full potential of the Platonic states.

\section{\label{sec:conclusion}Conclusion}
In conclusion, we demonstrated a scheme for the creation of the tetrahedron state, a 4-photon polarization state optimal in quantum mechanics for the estimation of polarization rotations. While a correctly aligned $N00N$ state can measure the rotation angle of a rotation with a known axis with Heisenberg scaling, one would need three $N00N$ state copies, each aligned with the different Cartesian axis to measure the three parameters of a spin rotation with the same scaling. This resource increase leads to a degradation of the asymptotic performance by a factor three in the appropriately weighted sum of the measured parameters' variances. The tetrahedron state, and other second-order unpolarized states, can be used to simultaneously measure all of the parameters of the rotation with Heisenberg scaling and are therefore not affected by this performance hit. Generalizations of these states tailored to transformations described with a minimum of $d$ parameters would yield $d$-fold improvements in the weighted sum of variances of said $d$ parameters. 
In this work, we created the tetrahedron state in the polarization of four photons by combining three photons themselves in a $N00N$ state attenuated along the horizontal axis with a horizontally-polarized single photon. Implementing this required interfering the output of two SPDC sources and a coherent state. The main source of noise in our experiment comes from events where more photons were created than those needed for the tetrahedron state. These events could be made less significant by reducing the optical losses, which in our experiment, could be achieved with with better sources and detectors. Furthermore, in principle, the rate of these insidious events can always be made proportionally arbitrarily small by decreasing the overall tetrahedron state creation rate and increasing the counting time accordingly. 
In part because of the low four-photon rate of our experiment, we were unable use the quantum state we created to measure parameters of a polarization rotation. Nevertheless, at the proof-of-concept level, we have created and characterized a state that shares many of the interesting qualitative features and symmetries of the optimal tetrahedron state. 
Progress in finding robust methods to create similar states in high-dimensional Hilbert spaces, both in optics and in other physical medium, could prove very impactful for many multiparameter metrology application like 3D-magnetometry, ellipsometry, and more. 

\begin{acknowledgments}
This work was supported by the Natural Sciences and Engineering Research Council (NSERC) of Canada, CIFAR, and the Austrian Science Fund (FWF), through BeyondC (F7113). It made use of some equipment purchased through grant FQXiRFP-1819 from the Foundational Questions Institute and Fetzer Franklin Fund, a donor-advised fund of Silicon Valley Community Foundation. HF, YBY, NLG, AOTP, and AMS wish to acknowledge acknowledge that the University of Toronto operates on the traditional land of the Huron-Wendat, the Seneca, and the Mississaugas of the Credit.
KB-F and AZG acknowledge that the NRC headquarters is located on the traditional unceded territory of the Algonquin Anishinaabe and Mohawk people. 
\end{acknowledgments}

\bibliography{biblio2}

\end{document}